\documentstyle[sprocl]{article}
\input{epsf}
\begin{document}

\title{INFALL OF A PARTICLE INTO A BLACK HOLE AS A MODEL FOR
GRAVITATIONAL RADIATION FROM THE GALACTIC CENTER}

\author{Carlos O. Lousto}

\address{
Department of Physics, University of Utah, Salt Lake City, UT 84112, USA}

\maketitle\abstracts{
I present here the results of the study of the gravitational radiation
generated by the infall (from rest at radius $r_0$)
of a point particle of mass $m_0$ into a Schwarzschild
black hole of mass $M$. We use Laplace's transform methods and find that
the spectra of radiation for $\sim5M<r_0<\infty$ presents a series of
evenly spaced bumps. The total radiated energy is not
monotonically decreasing with $r_0$, but presents a {\it joroba}
(hunch-back) at around $r_0\approx4.5M$.
I finally discuss the detectability of the
gravitational radiation coming from the black hole in the center of our
galaxy.}

\section{Perturbative approach}

Here I will report on work made in collaboration
with R. Price (see Ref.\ 1 for further details.)
The problem of a particle falling into a non-rotating
black hole can be treated in the regime where the particle contributes
perturbatively to the Schwarzschild metric. Thus, it was not surprising
that soon after Zerilli (1970) wrote down his equation describing the
propagation of gravitational waves on the Schwarzschild background, the
case of a particle falling from infinity, both at rest and with a finite
velocity, was solved using Fourier transform techniques.
The key to the resolution to the problem of the infall from a finite
distance (that had to wait 25 years) is the use of the Laplace's transform
method instead. In this case, Zerilli's equation reads
\begin{equation}\label{basiceq}
  \frac{\partial^2\Psi_\ell}{ \partial r*^2}
  +\left[\omega^2-V_\ell(r)\right]\Psi_\ell=-\dot{\psi}_0(r)
+i\omega\psi_0(r)+S(r,\omega)\ ,
\end{equation}
where the $\ell$-multipole of the waveform is related to the metric
components (in the Regge--Wheeler gauge) by
\begin{equation}\label{psidef}
  \psi_\ell(r,t)=\frac{r}{{\lambda}+1}\left[
    K+\frac{r-2M}{{\lambda}r+3M}\left\{ H_2-r\partial K/\partial r
    \right\} \right]\ ,
\end{equation}
where ${\lambda}\dot=(\ell+2)(\ell-1)/2\ .$ An important ingredient in
Eq.\ (\ref{basiceq}) is the source term $S(r,\omega)$ that we had to 
compute for the particle infalling along a geodesic. We also had to consider
the initial waveform $\psi_0(r)$ corresponding to a solution of the 
hamiltonian constraint. Here we have taken the particle limit of the
Brill-Lindquist initial solution (the results do not differ notably
had we chosen the Misner initial solution), and a time-symmetric
situation ($\dot{\psi}_0(r)=0$).

\section{Results}

To solve Eq.\ (\ref{basiceq}) we use the Green's function method, integrating
over the total effective source, i.e. the right hand side of
Eq.\ (\ref{basiceq}), to obtain the complex amplitude $A(\omega)$ of the
outgoing radiation. The main results we have obtained can be summarized
in the three figures we give here:
\begin{figure}
\centering\epsfxsize=280pt\epsfbox{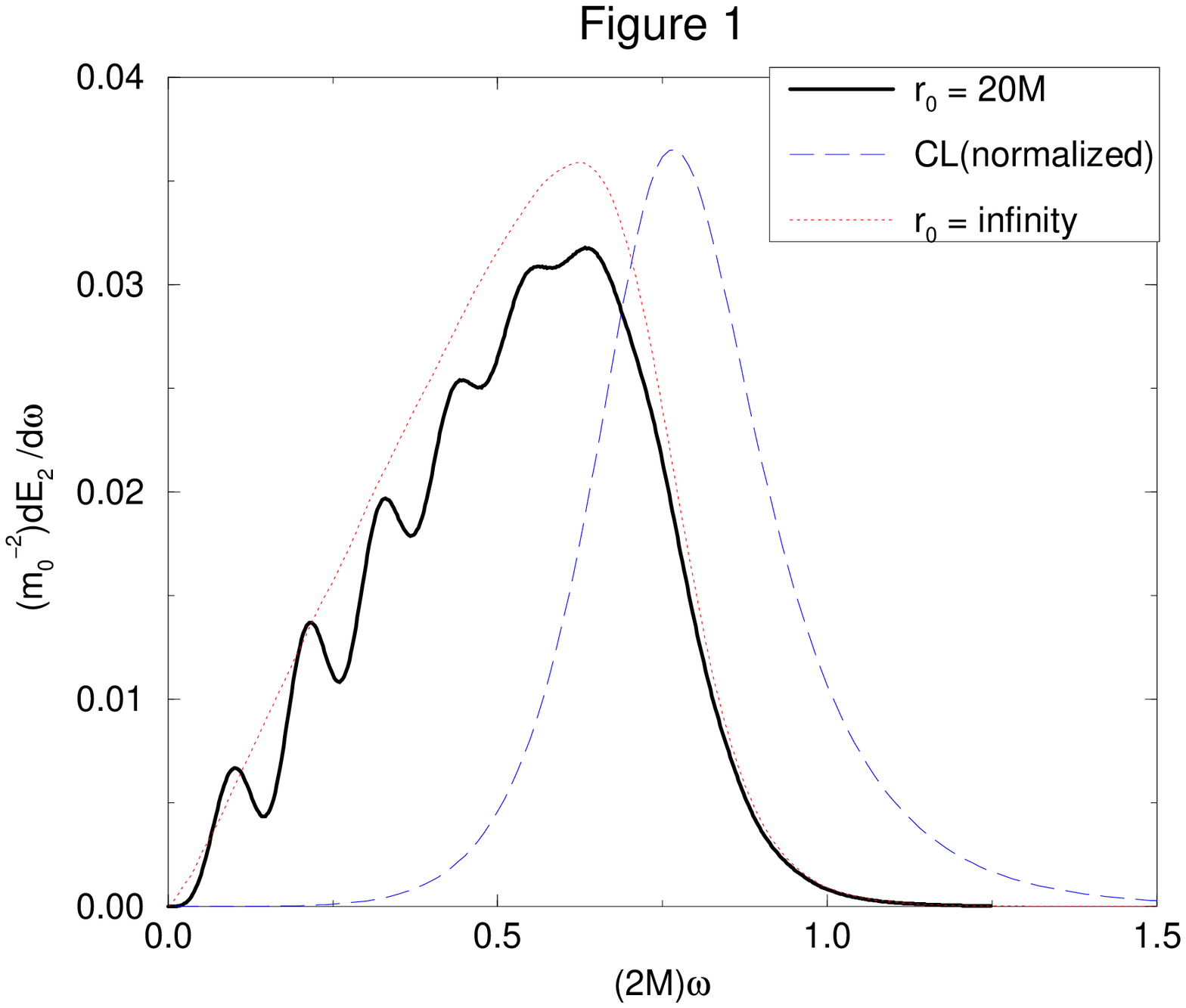}
\end{figure}
In Fig.\ 1 the new feature is the appearance
of bumps in the spectrum whose spacing decreases as $r_0$ increases. This
can be understood as a consequence of the interference between the initial
burst of radiation, soon after $t=0$, and the final one due to the
infall of the particle into the black hole,
\begin{figure}
\centering\epsfxsize=280pt\epsfbox{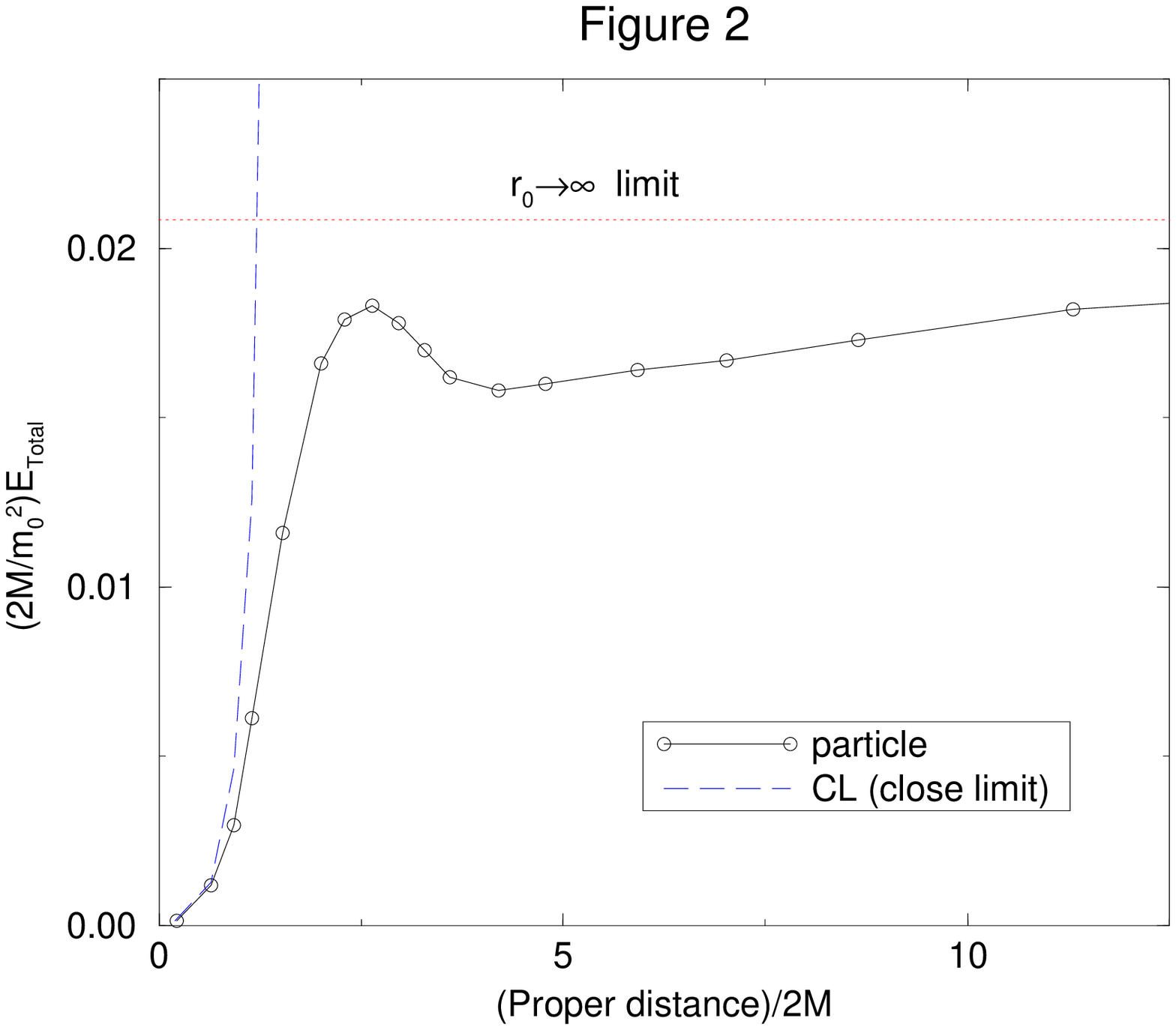}
\end{figure}
In Fig.\ 2
the novelty is the {\it joroba}, with a local maximum at $r_0=4.5M$. This
purely general relativistic effect can be attributed to the higher efficiency
in the generation of radiation by the initial burst close
to the maximum of the potential (located at $r\approx3.1M$.)
Here is a small challenge for supercomputers: To check up to which mass 
ratio the {\it joroba} survives in the nonlinear regime.
Fig.\ 3 shows the characteristic waveforms for different trajectories of
the infalling particle. The main observation here is that the early
$(u=t-r^*<0)$ behavior of the waveform depends only on the initial data
of the particle; while the late $(u>0)$
behavior only depends on the black hole characteristics
(quasinormal ringing). Here is the challenge for gravitational wave
detectors: To have enough sensitivity to resolve this two parts of
the waveform and then learn on both, the initial state of motion of
the particle and the central black hole.
\begin{figure}
\centering\epsfxsize=280pt\epsfbox{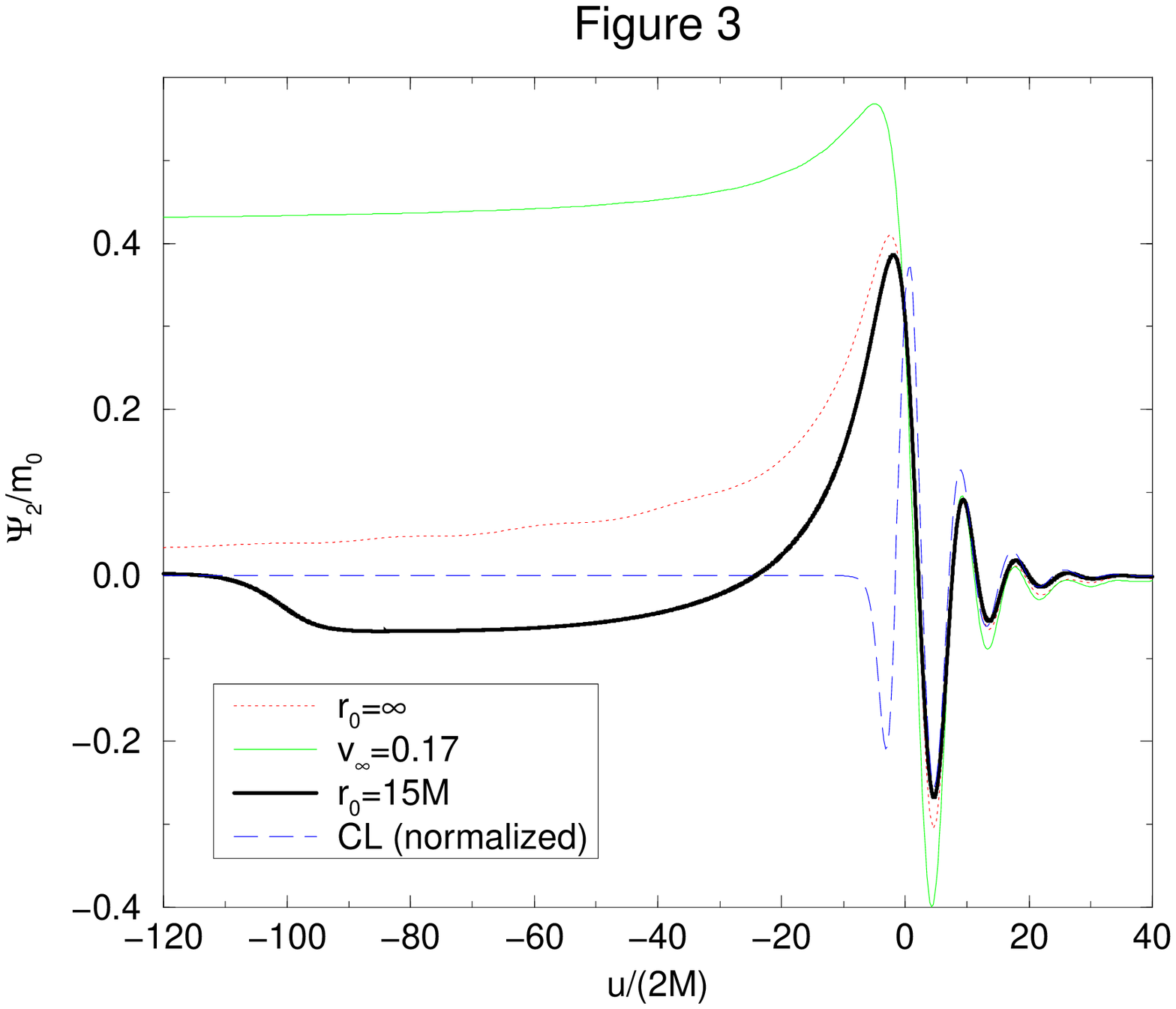}
\end{figure}

\section{Observational estimates}

The confirmation\cite{EG96} of the presence of a black hole with
$M=2.4\times10^6M_\odot$ in the center of our galaxy provides
an astrophysical scenario for testing our computations.
Let us first consider the amplitude of the metric perturbations as
the gravitational radiation reaches the earth. From the
waveforms given in Fig.\ 3 and Eq.\ (\ref{psidef}),
we find that $\Delta K\sim h\sim 10^{-17}(m_0/M_\odot)(8kpc/r)$, independent
of $M$, the mass of the black hole\footnote{Rotation of the hole and
angular momentum of the infalling particle may increase this value in
two orders of magnitude}. The duration of the final burst
of radiation can be estimated as $\Delta t\approx 3\times 10^{-4}(M/M_\odot)
sec\approx 12min$.
The next important issue is the frequency of such events. As an order of
magnitude estimate we can study the quantity $f=\rho\sigma\nu$, where
$\rho\approx6.5\times10^9M_\odot/pc^3$, is the density of stars near the
hole, $\sigma_{\rm cap}=16\pi M/v^2$ is the capture cross section, and
$\nu\approx400km/sec\approx v$
is the mean velocity of the stars relative to the hole. All this
together gives $f=7.5\times10^{-24}(M_\odot/m_0)(M/M_\odot)^2(c/v)sec^{-1}
\approx1\,{\rm event}/yr$.
Finally, from the spectrum of gravitational radiation we see that its
maximum takes place at a frequency $\omega_{\rm max}\approx7\times10^{4}
(M_\odot/M){\rm Hz}\approx0.03{\rm Hz}$.
With all this numbers\footnote{The tidal
forces of the black hole will disrupt the upper atmosphere of the infalling
star (generating a burst of electromagnetic radiation accompanying the
gravitational one), but leave its core (where most of the star's mass
resides) practically untouched. This justifies our particle approximation.}
in hand we conclude that the galactic center may well be among the
first positively detected sources of gravitational radiation (by LISA).
\section*{Acknowledgments}
I would like to thank A. Giazotto, 
J. Horvath, and R. Price for discussions on this problem, and NSF Grant
No. PHY0507719 for financial support.

\end{document}